\documentclass[oneside,11pt,a4paper]{report}
\usepackage[english, titlepage]{ku-frontpage}
\usepackage[utf8]{inputenc}
\usepackage{hyperref}
\usepackage[numbers,sort]{natbib}
\usepackage{pdfpages}

\assignment{PhD Thesis}
\author{Christian Hansen}

\usepackage[titles]{tocloft}
\title{Sequential Modelling with Applications to Music Recommendation, Fact-Checking, and Speed Reading}

\subtitle{} 
\date{Handed in: April 30, 2021\\ This thesis has been submitted to the PhD School of The Faculty of Science, University of Copenhagen}
\advisor{Advisors: Stephen Alstrup, Christina Lioma, Jakob Grue Simonsen}

\begin{document}
\maketitle
\pagenumbering{roman}


\phantomsection
\addcontentsline{toc}{chapter}{Abstract}\section*{Abstract}
Sequential modelling entails making sense of sequential data, which naturally occurs in a wide array of domains. One example is systems that interact with users, log user actions and behaviour, and make recommendations of items of potential interest to users on the basis of their previous interactions. In such cases, the sequential order of user interactions is often indicative of what the user is interested in next.
Similarly, for systems that automatically infer the semantics of text, capturing the sequential order of words in a sentence is essential, as even a slight re-ordering could significantly alter its original meaning. This thesis makes methodological contributions and new investigations of sequential modelling for the specific application areas of systems that recommend music tracks to listeners and systems that process text semantics in order to automatically fact-check claims, or "speed read" text for efficient further classification.

For music recommendation, we make three contributions: Firstly, a study of how the complexity of sequential music recommender methods relates to the diversity and relevance of the recommendations, and how diversification of recommendations can be used to control this trade-off. Secondly, we investigate how listening context impacts music consumption, which we use to motivate a new way of representing user profiles that captures sequential and contextual deviations from the user's typical music preferences. Thirdly, we improve the prediction of music skip behaviour in a listening session based on past skips.

For fact-checking, we make three contributions: Firstly, we construct the currently largest benchmark dataset of naturally occurring claims for training automatic fact-checking models. Secondly, we link and use eye-tracking data of humans reading news headlines to automatic fact-checking predictions. Thirdly, we present two models for detecting check-worthy sentences for fact-checking, which by the use of weak supervision and contrastive ranking, make steps towards better model generalization in a domain with very limited training data.

Lastly, for speed reading, we contribute a new model that utilizes the inherent punctuation structure of text for learning how to ignore a large number of words, while being equally or more effective than processing every word in the text.

\clearpage

\phantomsection
\addcontentsline{toc}{chapter}{Dansk Resumé}\section*{Dansk Resumé}
Sekventiel modellering indebærer at skabe mening i sekventielle data, som naturligt forekommer i en bred vifte af domæner. Et eksempel er systemer, der interagerer med brugerne, logger brugerhandlinger og adfærd, og giver anbefalinger af potentiel interesse for brugerne på baggrund af deres tidligere interaktioner. I sådanne tilfælde er den sekventielle rækkefølge af brugerinteraktionerne ofte en indikation af, hvad brugeren i fremtiden er interesseret i.
Tilsvarende er det vigtigt for systemer, der automatisk udleder semantikken i tekst, at repræsentere rækkefølgen af ord i en sætning, da selv en lille omstilling kan ændre dens oprindelige betydning væsentligt. Denne afhandling yder metodiske bidrag og nye undersøgelser af sekventiel modellering til specifikke anvendelsesområder for systemer der anbefaler musiknumre til lyttere, og systemer der behandler tekstsemantik for automatisk at foretage faktakontrol af udsagn eller "speed-read" tekst for effektiv klassificering.

Til musikanbefaling yder vi tre bidrag: For det første en undersøgelse af, hvordan kompleksiteten af sekventielle musikanbefalingsmetoder er relateret til diversitet og relevansen af anbefalingerne, og hvordan diversificering af anbefalinger kan bruges til at kontrollere dette trade-off. For det andet undersøger vi, hvordan lyttekontekst påvirker musikforbruget, som vi bruger til at motivere en ny måde at repræsentere brugerprofiler, der fanger sekventielle og kontekstuelle afvigelser fra brugerens typiske musikpræferencer. For det tredje forbedrer vi forudsigelsen af hvor brugerer skipper musiknumre i en session baseret på tidligere skip.

Til faktakontrol yder vi tre bidrag: For det første konstruerer vi det i øjeblikket største benchmarkdatasæt med naturligt forekommende udsagn til træning af automatiske faktakontrolmodeller. For det andet forbinder vi og bruger eye-tracking data fra mennesker, der læser nyhedsoverskrifter til automatiske faktakontrolforudsigelser. For det tredje præsenterer vi to modeller til detektering af kontrolværdige sætninger til faktakontrol, som ved hjælp af weak supervision og kontrastiv ranking tager skridt mod bedre model generalisering i et domæne med meget begrænsede træningsdata.

Endelig bidrager vi til speed reading problemet med en ny model, der udnytter tekstens tegnsætningsstruktur til at lære at ignorere et stort antal ord, samtidig med at den er mindst lige så god som hvis hvert ord i teksten bliver processeret.
\clearpage

\phantomsection
\addcontentsline{toc}{chapter}{Acknowledgements}\section*{Acknowledgements}
First and foremost, I am grateful to my three supervisors, Christina, Jakob, and Stephen, who have provided me with many lessons on both academia and science as a whole, as well as personal lessons I will keep forever. My weekly meetings with Christina and Jakob have always been a source of joy, as they were both scientifically enlightening but also simply entertaining. I would also like to thank my fellow PhD students, with whom I have worked during the years, Lucas, Dongsheng, Stephan, and Niklas.

During my PhD, I have worked together with great people at Edulab to improve how mathematics is taught in Denmark. This has been an excellent collaboration where real-world problems could meet academia with exciting results. Of all the people at Edulab, I would especially like to thank Lotte, Kasper, Morten, Jakob, Stani, and Daniel.

I was fortunate to spend a summer away from Denmark as an intern at Spotify Research in London. I got to work with many wonderful people, where I would especially like to thank Rishabh, Brian, Federico, Lucas, and Mounia.

I also want to thank Rastin and Benjamin for a great collaboration at Danmarks Nationalbank. I realized how lucky I am that papers in computer science are so much shorter than in economics. 

From my personal life, I would first like to thank all my friends and family who have put up with my wacky work schedule during the last few years. Especially I would like to thank Mikkel, Helena, and David, who have been a great support. Lastly, I would like to thank the two most important people during my PhD, my twin brother Casper and my fiancée Spring. Casper has been a great source of inspiration in our work together and has been the best collaborator anyone could ever hope to work with. Spring has been a constant source of support and love, who not only accepts that I like to talk about my work, but also perfectly feigns interest in it.

\tableofcontents
\clearpage
\pagenumbering{arabic}

\chapter{Introduction}
Many domains have a natural sequential ordering of their data, where ignoring the sequential ordering risks throwing away valuable information. 
One such domain is that of recommender systems, where a user's current preferences may be related to both recent interactions, characterising their current needs, and also long term behavior, expressing general preference patterns. 
In both the long and short term, how users interact with items happens in a sequential order, which typically implies a sequential dependency between the interactions. 
Examples of such sequential dependencies could be in e-commerce systems, where a user might be looking for clothes matching another recent purchase, or on music streaming platforms, where a user's current preference may be learned over a larger number of previous interactions.
A seemingly different, but in fact highly analogous, domain is that of semantic inference of text, where the sequential word order is paramount for understanding meaning correctly, e.g., ''\textit{The lion ate a chicken}'' or ''\textit{The chicken ate a lion}''. This is broadly known as \textit{term dependence} and has been long investigated by both linguists and computer scientists \cite{raganato-etal-2017-neural}.

Utilizing and learning from sequential data is the topic of sequential modelling. In this thesis, we focus on music recommendation and textual fact-checking as two specific application areas of sequential modelling. We describe these below.

\subsection*{Music Recommendation}
Recommender systems are integral in helping users navigate through the potentially huge number of items available on a content platform. They do so by presenting users with relevant items that match their preferences. Training such preference models is a challenge as the available training data is very sparse. This happens because a single user normally only interacts with a small fraction of the available items \cite{guo2012resolving}. In this setup, two distinct ideas are used for learning user preferences: Firstly, similar users tend to have similar preferences, which can guide recommendation. Secondly, a user's preferences may be based on certain item properties, such that users can be recommended items similar to what they have previously enjoyed.
These are the core ideas behind collaborative filtering \cite{koren2009matrix} and content-based filtering \cite{van2013deep, pazzani2007content}, respectively.
To combine the benefits of both, many modern recommender systems are hybrid approaches utilizing both the item features and shared preferences \cite{beutel2018latent, dai2016deep, ren2019repeatnet}. 

Music recommendation has certain particularities different from the recommendation of other items, such as movies or books \cite{schedl2018current}. Consumption of music is highly sequential, as a single session easily consists of listening to tens of tracks, where the likelihood of a user enjoying a track is not only dependent on the track itself, but also on how it appears in the sequence \cite{schedl2018current}. Users also tend to repeat previously listened tracks \cite{anderson2014dynamics}, and have their current music preference strongly influenced by situational and contextual aspects \cite{cebrian2010music, kaminskas2013location, pettijohn2010music}. Additionally, music listening is often a passive activity where the tracks are recommended and played automatically. This means that users have to actively skip a track to elicit a negative feedback signal. In contrast, not skipping a track cannot necessarily be interpreted as a positive feedback signal, as the user may simply not be actively engaged.

In this thesis, we incorporate and exploit the above mentioned particularities to design new methods for recommender systems. 
Firstly, we investigate how the complexity of the ranker used by the recommender system is related to both the relevance and diversity of the recommendations, where both sequential and non-sequential models are considered. To increase recommendation diversity, different diversification methods are investigated to explore their relevance and diversity trade-offs.
Secondly, we perform a large-scale study on how context impacts music consumption, which we use to motivate a novel approach for generating dynamic user embeddings capturing sequential and contextual deviations from the user's typical music preferences.
Thirdly, we investigate to what extent a user's skip behaviour can be predicted as a sequential classification task, where the first half of a session is used to predict the skip behavior of the second half, with the aim of better understanding the difficulties of modelling skips.

\subsection*{Fact-Checking}
Misinformation is spreading at increasing rates, where especially fake news has a tendency to reach a larger audience and spread faster compared to news that is factually true \cite{vosoughi2018spread, zubiaga2018detection}. This spread of misinformation has long been considered a highly pressing societal issue by the World Economic Forum \cite{howell2013}, and due to the scale of the issue, automatic solutions for fact-checking are necessary. An automated fact-checking pipeline \cite{thorne2018automated} normally consists of three steps: (i) selecting check-worthy sentences, which are sentences containing a claim worth fact-checking, (ii) gathering related evidential information to those sentences which can help decide factuality, and (iii) using the evidence to infer the factuality of each check-worthy sentence. We describe each step below.

For step (i), the automatic selection of check-worthy sentences is considered a ranking task, which for a given text, e.g., transcribed political speeches and debates, aims to provide a ranked list of sentences in the order of how relevant they are to fact-check. A sentence is said to be check-worthy if it contains a factual claim that is of interest to determine the factuality of \cite{hassan2017toward}. This means that even though a sentence like ''\textit{My dog is brown}'' is factual, it is not check-worthy as its factuality would be of little interest to most people.
Step (ii) and (iii) are typically considered jointly, as the modelling approach is determined by the structure of the input. Multiple different sources for obtaining evidential information have been considered for determining the factuality of a claim.
This includes querying search engines with the claim as a query \cite{karadzhov2017fully}, or limiting the search to specific domains such as social media posts on Twitter \cite{ba2016vera}. Knowledge graphs have also been used for extracting facts related to the central entities of a claim \cite{ciampaglia2015computational}. Lastly, previously fact-checked claims can also be used as evidence when fact-checking new claims \cite{shaar2020known}, as semantically similar claims may share the same factuality.

In this thesis, we make a series of contributions toward all three steps of the fact-checking pipeline. Firstly, we construct the largest dataset of naturally occurring claims. We do this by crawling claims from 26 fact-checking websites, where associated evidence obtained from a search engine and rich metadata is made available. We verify the dataset's usefulness for fact-checking by ablating the effectiveness improvement of including both the evidence and metadata. Secondly, we explore other modalities of fact-checking evidence, where we study to what extent eye-tracked data from users can be used for inferring factuality. Thirdly, we propose new check-worthiness models utilizing weak supervision and contrastive ranking to make more accurate predictions.
Lastly, unrelated to the fact-checking pipeline, but related to the sequential modelling used in our proposed models, we consider the task of speed reading. Speed reading is the task of processing as few sequential inputs as possible without compromising model effectiveness. We propose a new model, which utilizes the inherent punctuation structure of text for learning how to ignore significant parts of the input sequence, while being equally or more effective than processing the entire sequence.

\section{Research Outline}\label{s:research-outline}
This thesis is composed of eight published articles, each of them presented as a separate chapter. These eight chapters are clustered into three themes, according to their domain of application. These three themes are: music recommendation, fact-checking, and speed reading. 

This section provides an overview of the primary research questions tackled in each thesis chapter, and how these were investigated. For each research question, the relevant background material and main findings are briefly covered.

\subsection{Music Recommendation}
\subsection*{Chapter \ref{chapter1}: Shifting Consumption towards Diverse Content on Music Streaming Platforms}

The meaning of diversity in recommendations is influenced by the domain and task \cite{kunaver2017diversity}, but a general definition is that diversity is the opposite of similarity among the recommended items \cite{bradley2001improving}. The aim of this work is to explore how diversity can be included in sequential music recommendation, where the user is passively listening or may choose to actively skip a given recommendation. The explicit active choice of skipping can be seen as interrupting the user experience, and is detrimental to the overall user satisfaction. Thus, when considering diversity in sequential recommendation, any recommendation given as a consequence of increasing the diversity should still be relevant to the user. This contrasts the work on list recommendation \cite{ashkan2015optimal, bedi2015novel, ribeiro2015multiobjective, wasilewski2016incorporating}, where the user is presented a number of items. In this case, if a subset of the recommended items are highly relevant, it is possible to include a selection of less relevant but more diverse items in the rest of the list.

Although there is a potential detrimental cost of increasing the diversity in sequential recommendation, there are a number of potential benefits which make diversity worth pursuing. The first one is that diversity can help users discover new interests \cite{lathia2010temporal,zhang2012auralist,ziegler2005improving}, which have been linked to long term user retention \cite{anderson2020algorithmic,DBLP:conf/www/spotify-www2021}. The second one is that diversity can help avoid the rich-get-richer problem \cite{salganik2006experimental}, where a small subset of items receive a large amount of the interest as a consequence of how the recommender systems are trained. 

The above leads to the following research questions:
\begin{description}
\item[RQ1] To what extent can diversity be included in sequential recommendation, and how can its effect on the relevance of the recommendations be controlled?
\item[RQ2] What is the relation between the complexity of the ranker used by the recommender system and the diversity of the
recommendations?
\end{description}
To answer \textbf{RQ1}, we first define two notions of diversity relevant to music recommendation. The first notion defines
diversity with respect to popularity,
and the second with respect to
personalization. We investigate four different methods for increasing diversity, and we empirically evaluate how they affect the relevance of the recommended items to the user. 

To answer \textbf{RQ2}, we evaluate four rankers of increasing complexity, where complexity of the ranker refers to the amount of user information and size of the model it uses for the recommendation. Specifically, we evaluate how each ranker fares with regards to both the effectiveness and diversity of the recommendations. 

Our findings regarding both \textbf{RQ1} and \textbf{RQ2} show that i) it is possible to increase the diversity of the recommended items with little to no decrease in relevance, while even higher diversity can be achieved if a larger decrease in relevance is accepted; and ii) as the complexity of the rankers increases, there is a tendency for the recommendations to get more relevant but less diverse.

\subsection*{Chapter \ref{chapter2}: Contextual and Sequential User Embeddings for Large-Scale Music Recommendation}



In large-scale settings, it can be beneficial from an efficiency perspective to express user-item relevance using a simple vector operation between a user and item embedding~\cite{bachrach2014speeding}, such as the cosine similarity. This contrasts the methods used for answering \textbf{RQ1-2}, where the relevance between a user and an item was estimated by more computationally expensive models. However, embedding-based methods are not limited to using a static user embedding, which corresponds to using the same user embedding for each session. Rather, how often the user embedding is updated is a trade-off between recommendation effectiveness and model efficiency. In this work, we consider the problem of learning user embeddings that are updated based on the sequential consumption of past sessions and the current context. 

Regarding consumption, prior work has shown that music consumption is highly driven by recency \cite{anderson2014dynamics}, as users tend to repeat the same tracks often \cite{anderson2014dynamics,chen2015will}. Regarding context, it has been established that the tracks a user listens to are often context-dependent, such as based on the time of the day \cite{cebrian2010music}, location \cite{kaminskas2013location}, weather \cite{pettijohn2010music}, and current season \cite{park2019global}. However, these studies on contextual dependency have been done on small datasets, and do not investigate its impact on recommendation effectiveness.

Motivated by the above, we ask the following research questions:
\begin{description}
\item[RQ3] To what degree does music consumption depend on context?
\item[RQ4] To what extent can sequential and context-dependent user embeddings better anticipate a user’s music consumption?
\end{description}
To answer \textbf{RQ3}, we explore historical data from an online music streaming service, containing information about the tracks streamed by a sample of 200,000 users over a two month period. In our analysis, we consider two types of contexts: the time of the day (temporal context), and the device used for music streaming (device context), such as mobile, desktop, speaker, etc. 

For both sessions associated with an individual user and across all users, we find that the tracks within a session are more similar to the tracks in sessions of the same context, compared to sessions from a different context. Furthermore, we also find that tracks deviating highly from a user's average preferences, represented by what tracks they usually listen to, are more likely to be skipped by the user.

The findings from \textbf{RQ3} support the idea of developing sequential and context-aware models for representing user preferences. To this end, we address \textbf{RQ4} by introducing a new recurrent model that generates user embeddings matching the user's preference based on their current context (in the current session) and from the sequence of past consumed sessions. Our model is trained to maximize the cosine similarity between the user embedding and tracks played during a given session. We find that a highly effective way to learn the user embedding is by fusing a global context-independent embedding (representing the user's average preferences) with a learned sequential and contextual offset embedding (representing the sequential and contextual deviations to the user's average preferences). We experimentally compare our model to state-of-the-art embedding-based models in a range of ranking tasks, where we observe improvements in ranking effectiveness upwards of 10\%. Interestingly, we find that the largest gains occur in the least frequent contexts, highlighting the model's ability to accurately learn the contextual deviations between sessions.

\subsection*{Chapter \ref{chapter3}: Modelling Sequential Music Track Skips using a Multi-RNN Approach}
In \textbf{RQ1-4} we considered the problem of music recommendation within or between sessions, with the aim to recommend music tracks that the user is unlikely to skip. To this end, understanding and inferring the skip behaviour of users within a session is very important, and was in fact a competition challenge for the WSDM Cup 2019 \cite{brost2019music}. In this challenge, a session was split in two halves, such that the task was to predict the individual track skips occurring in the last half of the session, based on the skips made by the user in the first half. For all tracks in the session, track features (e.g., popularity or musical features like strength or flatness) were available. The only difference between the two halves was the existence of user feedback in the form of skips. 
This problem shares similarities with the general problem of sequence-to-sequence prediction \cite{sutskever2014sequence}, but differs from it in that only the user feedback is unknown; whereas, in typical sequence-to-sequence problems, the whole predicted sequence is unknown. Interestingly, it has been observed that the skip behaviour of users is not entirely dependent on the actual track, but also largely depends on whether the user skipped the previous track \cite{brost2019music}. 

Motivated by the above, we raise the following research question:
\begin{description}
\item[RQ5] To what extent is future sequential skip behaviour predictable by the past?
\end{description}
To answer \textbf{RQ5}, we propose an encoder-decoder model based on two distinct stacked recurrent neural networks (RNNs) using long short-term memory (LSTM) units. This encoder-decoder architecture is a type of neural architecture that is often used for sequence-to-sequence problems, such as machine translation \cite{gehring2017convolutional, wu2016google}. 

Our model was the second best performing model in the WSDM Cup 2019 competition (out of 45 teams), with the best performing model \cite{zhu2019session} also being based on a similar encoder-decoder architecture. We investigated how the model accuracy differs between predicting whether the first track was skipped compared to the average accuracy across all tracks in the second half of the session. We observed that it was notably easier to predict for the first track (0.807 accuracy) compared to the whole second half (0.641 mean average accuracy). This highlights the difficulty of predicting the skip behaviour far out in the future, as if otherwise the accuracy of the two settings should not be drastically different.

\subsection{Fact-Checking}
\subsection*{Chapter \ref{chapter4}: MultiFC: A Real-World Multi-Domain Dataset for Evidence-Based Fact Checking of Claims}
Automatic fact-checking is the task of predicting the factuality of a claim, typically based on associated evidence and metadata. The evidence is often automatically collected from external knowledge sources, such as search snippets from a search engine \cite{thorne2018automated, allein2020time, popat2018declare}. However, existing datasets consist of either a small amount of naturally occurring claims \cite{mihalcea2009lie, zubiaga2016analysing} or artificially constructed claims \cite{thorne2018fever}. 

Motivated by the above, we pose the following research question:
\begin{description}
\item[RQ6] How can a large dataset of real-life claims with accompanying evidence be created, to aid in the research of automatic fact-checking?
\end{description}
To answer \textbf{RQ6}, we built automatic crawlers for 26 active fact-checking websites.\footnote{\url{https://reporterslab.org/fact-checking/}} From each website, the crawlers automatically extract the claim, its associated factuality label, and any accompanying metadata that is made available by the individual websites (e.g., tags, speaker name, and publication date). The total crawling resulted in a dataset consisting of 34,918 claims and it was the largest dataset of its kind at the time of publication. To enrich the dataset with evidence, we used the claims verbatim as queries to the Google search API, from which we crawled the top ten retrieved results. 
To verify the usefulness of the dataset, we trained a state-of-the-art fact-checking model, and ablated the effectiveness impact of using only the claim, including search snippets as evidence, as well as metadata. We found that both evidence and metadata are beneficial for improving the effectiveness of the model.

\subsection*{Chapter \ref{chapter5}: Factuality Checking in News Headlines with Eye Tracking}
In \textbf{RQ6} fact-checking was done using claims, metadata, and associated evidence extracted from the web. In this work, we investigate other modalities of evidence, which can be used to determine the factuality of a claim. Specifically, we consider whether data from eye-tracking can be used to infer the factuality of a claim, as eye-tracking has previously been used in information retrieval to infer relevance~\cite{AjankiHKPS09,Buscher:2012:ADE:2070719.2070722,BuscherDE08,Hardoon:article,LobodaBB11,PuolamakiAK08}.
Additionally, eye-tracking has been used to investigate how users engage with news content, where it has been observed that users tend to read false news faster~\cite{Gwizdka14}, as well as putting more visual attention on credible news posts \cite{doi:10.1177/1461444818791520}. These observations establish a relation between a person's reading behaviour and the factuality and credibility of the read material. 

Motivated by the above, and focusing on the domain of news, we raise the following research question:
\begin{description}
\item[RQ7] To what extent can the factuality of a news headline be inferred using only eye-tracked data?
\end{description}
To answer \textbf{RQ7}, we conducted a user study where the participants were eye-tracked while reading news headlines that are either true or false. The headlines were crawled from a reputable local newspaper and a subset of these were manually falsified using a set of consistent semantic transformations. The participants were eye-tracked while reading the headlines, and five different measures were collected for each headline: the total gaze duration, total fixation duration, total fixation count, average fixation duration, and first fixation duration. Fixation corresponds to a stable eye position within a dispersion threshold, above a duration threshold, and gaze is the cumulative duration of a sequence of consecutive fixations. 
For inferring the factuality of a headline,  we proposed an ensemble model that combines the average factuality prediction of a set of participants to produce the final prediction.
The prediction for each participant was modelled as an average of two simple second order logistic models. We chose simple models due to the low amount of available data. We found that the ensemble model over all the participants obtained a mean AUC of 0.69, whereas using only a single participant led to an AUC of 0.55. Thus, it is possible to infer the factuality of a news headline using only eye-tracked data, but the effectiveness is highly dependent on having data from multiple people.

\subsection*{Chapter \ref{chapter6}: Neural Check-Worthiness Ranking with Weak Supervision: Finding Sentences for Fact-Checking}
Automatic fact-checking methods are trained on claims typically deemed interesting by (reputable) news sources or fact-checking websites, and are as such often manually selected for further fact-checking. The task of check-worthiness prediction is to develop automatic methods for filtering texts, e.g., transcribed political speeches and debates, by assigning a score to each sentence \cite{hassan2017toward}. This score aims to reflect the degree to which a sentence requires fact-checking. 

Most existing research, at the time of publication of this article, had focused on using hand-crafted features to predict check-worthiness, such as bag-of-words representations, sentiment scores, and embedding averages \cite{gencheva2017context,hassan2017toward,jaradat2018claimrank,patwari2017tathya}, rather than representation learning approaches using recurrent neural networks or transformers. Based on the learned check-worthiness scores, a ranked list can be generated for prioritizing which sentences should be fact-checked.

The choice of using models based on hand-crafted features may be due to the limited training data available~\cite{nakov2018overview}, as more complex models may be more prone to overfitting if used with limited training data. Check-Worthiness is a domain with high availability of data, but small amounts of labelled data, and for such domains weak labelling has been used successfully \cite{dehghani2017neural,zamani2018neural,nie2018multi}. Weak labelling is the process of using an existing classifier to get low-quality labels, also known as weak labels, on a typically large amount of currently unlabelled data. The main idea is that these weak labels can then be used as training data, to improve model generalizability. 

Motivated by the above, we ask the following research question:
\begin{description}
\item[RQ8] Can weak supervision be used for making check-worthiness predictions more accurate?
\end{description}
To answer \textbf{RQ8}, we extract a large number of sentences originating from political speeches and debates from the American Presidency Project.\footnote{\url{https://web.archive.org/web/20170606011755/http://www.presidency.ucsb.edu/}} These are labelled using ClaimBuster \cite{hassan2017toward}, an existing check-worthiness method with a publicly available API.\footnote{\url{https://idir.uta.edu/claimbuster/api/}} We use a recurrent neural network model that is pretrained on the weakly labelled data, which we evaluate with and without the weakly labelled data. We experimentally show that our model is more effective than state-of-the-art baselines, and that using the weakly labelled data significantly improves effectiveness. While our model greatly outperforms the weak labeller, the only baselines benefiting from the weakly labelled data do not perform better.

\subsection*{Chapter \ref{chapter7}: Fact Check-Worthiness Detection with Contrastive Ranking}

Existing methods for check-worthiness prediction are trained as a classification task \cite{gencheva2017context,hassan2017toward,jaradat2018claimrank,patwari2017tathya,DBLP:conf/www/hansen-www2019-factcheck,wright2020claim}, even though they are typically evaluated as a ranking task. As an extension to the model proposed for answering \textbf{RQ8} in the previous chapter, we consider how a ranking objective could be incorporated during training, as formulated by the following research question:
\begin{description}
\item[RQ9] How can ranking be part of training check-worthiness prediction models?
\end{description}
To answer \textbf{RQ9}, we are motivated by the finding in previous work showing a large term overlap between claims and non-claims \cite{le2016towards}. Because of this, we posit that check-worthiness models may face difficulties differentiating between highly similar sentences with opposing labels. To this end, for each sentence in our dataset, we find the nearest semantically similar sentences with opposing labels, denoted as contrastive sentences, and we use these as a set of tuples for training. In addition to the standard cross entropy classification loss of our model, we extend it with a hinge ranking loss that better learns to separate the constrastive sentences. We experimentally validate that including the ranking objective on contrastive sentences significantly improves ranking effectiveness, compared to our previous model.

\subsection{Speed Reading}
\subsection*{Chapter \ref{chapter8}: Neural Speed Reading with Structural-Jump-LSTM}

For the recommendation and fact-checking problems considered so far, we have proposed sequential models for inference and representation learning. Common to these models is that they all consist of recurrent neural networks for processing a sequence of inputs in its entirety. We now consider the problem of whether it is necessary to process every input, or whether parts can be ignored without compromising effectiveness. This has been explored in what is called ''speed reading'', which is based on solving text-based tasks using sequential models with the additional goal of making as few state updates as possible in the recurrent model. 

Speed reading tasks have traditionally been solved by two types of models. The first type is jump-based models \cite{skimmingGoogle, fu18, yu2018fast}, which during reading can choose to jump a certain number of steps ahead in the sequence, or terminate the reading of the sequence entirely when enough information is obtained to solve the task. 
The second type is skip (or skim) based models \cite{seo2017neural, campos2017skip}, which in addition to a full state update, can choose to either ignore the current input, thereby not making any state updates, or to skim the input and make a reduced state update. Common to all models is that they make the decision to ignore part of the sequential input based on the current and previous inputs, but do not utilize any inherent structure in the sequence. 

Our motivation is that, in sequences such as text, punctuation is a type of inherent structure that humans use to guide our reading behaviour. Therefore, punctuation  could potentially be utilized to determine how the jumps could be done in speed reading text. Inspired by this, we raise the following research question:
\begin{description}
\item[RQ10] To what extent can inherent text structure be used for defining the jumps in a speed reading model?
\end{description}
To answer \textbf{RQ10}, we propose a new hybrid speed reading model, Structural-Jump-LSTM, which combines both jumping and skipping of an input. The jumping is based on exploiting the punctuation structure, such that a jump is made towards either a comma, the end of a sentence (.!?), or the end of the document. We evaluate our model empirically in text classification and question-answering, and compare it against state-of-the-art speed reading models. We find that our model obtains the overall lowest number of state updates, corresponding to processing the fewest number of sequential inputs. Additionally, we find that speed reading models can often produce more accurate predictions than processing the entire sequence (i.e., full text), due to better generalization, which has similarly been observed in related work \cite{skimmingGoogle, seo2017neural}.

\section{Summary of Contributions}
This thesis makes a number of contributions for sequential problems faced in music recommendation, fact-checking, and speed reading. We summarize the contributions below: 
\begin{itemize}
    \item The first contribution is a study on diversity in sequential recommendation, using two notions of diversity related to popularity and user personalization. To this end, we first propose and evaluate multiple rankers of increasing complexity to study how their complexity impacts the diversity and relevance of the recommendations. Next, to increase the diversity, we investigate different diversification methods to explore their trade-off between increasing diversity and potentially reducing relevance. We find that rankers of high complexity result in more accurate but less diverse recommendations, while diversification methods enable increasing the diversity with little to no reduction in relevance.
    
    \item The second contribution is a study on the impact of context on music consumption, where we find that tracks within a listening session are most similar to the tracks from sessions in the same context. Motivated by this, we propose a new sequential model for dynamically generating user embeddings adapting to the contextual deviations from a user's general music preferences. Compared to state-of-the-art embedding-based baselines, we find modelling the contextual deviations to be effective, as seen by ranking improvements of upwards of 10\% in a range of ranking tasks.
    
    \item The third contribution is an investigation of the extent to which a user's sequential skip behaviour in a listening session is predictable by past skips. To this end, we propose an encoder-decoder model for predicting future unknown skips in a sequence of known recommended tracks. We show that as less recent skip information is available, the accuracy drops significantly, highlighting that the recommended track is not the only factor affecting the act of skipping.
    
    \item The fourth contribution is the construction of the largest-to-date fact-checking dataset of naturally occurring claims crawled from 26 active fact-checking websites. The claims are accompanied by evidence pages retrieved from a search engine, using the claims as queries, as well as rich metadata. We experimentally highlight the benefits of utilizing both the evidence and metadata, as seen by their impact on improving effectiveness.
    
    \item The fifth contribution is a study of how well factuality of a headline can be determined exclusively using eye-tracking data. The eye-tracking data was obtained from a user study where the participants were eye-tracked while reading factually true and false news headlines. We find that when eye-tracking data was pooled from multiple participants using an ensemble approach, factuality could be reasonably predicted with an AUC of 0.69, highlighting that eye-tracking can be used as a new modality for fact-checking methods.

    \item The sixth contribution is a new model for detecting check-worthy sentences for fact-checking. We find that training neural models in this domain is heavily limited by small amounts of training data, to which end we propose a strategy for using weak supervision, which significantly improves effectiveness.
    
    \item The seventh contribution is an improved model for detecting check-worthy sentences. Motivated by the observation of a large term overlap between claims and non-claims, which are highly similar to check-worthy and non-check-worthy sentences, we propose a model with a ranking-based objective, that better separates sentences with high semantic overlap, but opposing labels.
    
    \item The eighth contribution is a new speed reading model, which utilizes the inherent punctuation structure of text for learning how to ignore significant parts of the input sequence, while being equally or more effective than processing the entire sequence.
\end{itemize}

\section{Future Work}
Based on the contributions presented in this thesis, we outline some potential directions for future work below.

\subsection*{User effort as a measure of quality for sequential recommendation}
When evaluating the recommendations of a given recommender system, we normally aim to optimize the relevance of the recommendations. For music recommendation, this means minimizing the number of skips done by a user. This measure of recommendation quality is potentially flawed, because not all skips require the same amount of user effort. We can imagine at least three scenarios: 
\begin{itemize}
    \item The listening device has the screen turned off and is simply used for playing music, in which case turning on the screen to skip a track takes a moderate amount of effort;
    \item The user is actively using the device but does not have the music player open. In this case skipping a track takes less effort than in the previous example;
    \item The user has just skipped a track, in which case an immediate subsequent skip would take very little effort.
\end{itemize}
Based on these scenarios, we posit that collecting basic information about the state of the listening device during a session would allow estimating an approximate effort level of a skip. Rather than training recommender systems to minimize the number of skips, an alternative task would be the minimization of user effort, which could potentially better correlate with user satisfaction in passive listening sessions.

\subsection*{Handling bias in sequential recommendation}
When training sequential recommender models we currently assume that relevance feedback is purely dependent on the recommended item, even though this assumption is partly violated by the sequential dependencies between the feedback signals. One such example of a sequential dependency is the finding that users are much more likely to skip if they have just made a previous skip \cite{brost2019music}. If these dependencies can be reliably modelled, they could be included for correction during training, which could lead to more effective recommender systems. This can be seen as a bias correction for sequential models comparable to the debiasing done for list-wise recommendation in unbiased learning to rank \cite{joachims2017unbiased}.

\subsection*{Efficient feedback-based re-ranking for sequential recommendation}
Deploying a sequential recommender system that incorporates immediate feedback from a user requires a re-ranking after each interaction. To accommodate this, research into highly efficient sequential recommender systems is required and worth investigating. For non-sequential recommender systems, very efficient hashing-based methods have been investigated, where users and items are represented as hash codes \cite{DBLP:conf/www/hansen-www2021-phd, DBLP:conf/sigir/hansen-sigir2020-coldstart}, which require very little storage and enable very fast distance computations. However, hashing-methods have so far not been investigated for the domain of sequential recommendation. We posit that this is a direction worthy of further investigation.

\subsection*{Is automatically collected evidence sufficient for determining the factuality of a claim?}
In this thesis, in the context of fact checking, we considered the usage of automatically collected evidence as returned by a search engine when using claims as queries. While utilizing this evidence significantly improves the effectiveness of the factuality prediction, compared to only using the claim, many claims still remain difficult to fact check correctly. However, it has not currently been investigated whether this difficulty is due to insufficient evidence, or lack of better modelling, or inherent difficulty of the claim itself. To this end, it would be interesting to perform a user study of how well human assessors are able to determine claim factuality using only the same evidence as used by the fact-checking model. Additionally, it would provide a gold standard of human performance in the setting of evidence-based fact-checking.

\newpage

\section{List of Publications}
The following published articles are included as chapters of this thesis ($^*$ denotes equal contribution):
\begin{itemize}
    \item Christian Hansen, Rishabh Mehrotra, Casper Hansen, Brian Brost, Lucas Maystre, Mounia Lalmas (2021). Shifting Consumption towards Diverse Content on Music Streaming Platforms. In WSDM, pages 238-246. \cite{wsdm-spotify}.
    \item Casper Hansen, Christian Hansen, Lucas Maystre, Rishabh Mehrotra, Brian Brost, Federico Tomasi, Mounia Lalmas (2020). Contextual and Sequential User Embeddings for Large-Scale Music Recommendation. In RecSys, pages 53-62. \cite{DBLP:conf/recsys/hansen-recsys-spotify}.
    \item Christian Hansen, Casper Hansen, Jakob Grue Simonsen, Stephen Alstrup, Christina Lioma (2019). Modelling Sequential Music Track Skips Using a Multi-RNN Approach. In WSDM Cup. \cite{DBLP:conf/www/hansen-wsdmcup-2019}.

    \item Isabelle Augenstein, Christina Lioma, Dongsheng Wang, Lucas Chaves Lima, Casper Hansen, Christian Hansen, Jakob Grue Simonsen (2019). MultiFC: A Real-World Multi-Domain Dataset for Evidence-Based Fact Checking of Claims. In EMNLP, pages 4685-4697. \cite{augenstein-etal-2019-multifc}.
    \item Christian Hansen, Casper Hansen, Jakob Grue Simonsen, Birger Larsen, Stephen Alstrup, Christina Lioma (2020). Factuality Checking in News Headlines with Eye Tracking. In SIGIR, pages 2013-2016. \cite{DBLP:conf/sigir/sigir2020-eyetracking}.
    \item Casper Hansen, Christian Hansen, Stephen Alstrup, Jakob Grue Simonsen, Christina Lioma (2019). Neural Check-Worthiness Ranking with Weak Supervision: Finding Sentences for Fact-Checking. In Companion Proceedings of WWW, pages 994-1000. \cite{DBLP:conf/www/hansen-www2019-factcheck}.
    \item Casper Hansen, Christian Hansen, Jakob Grue Simonsen, Christina Lioma (2020). Fact Check-Worthiness Detection with Contrastive Ranking. In CLEF, pages 124-130. \cite{DBLP:conf/clef/hansen-clef-2020}.

    \item Christian Hansen, Casper Hansen, Stephen Alstrup, Jakob Grue Simonsen, Christina Lioma (2019). Neural Speed Reading with Structural-Jump-LSTM. In ICLR. \cite{DBLP:conf/iclr/hansen-iclr}.
\end{itemize}
Furthermore, in addition to the research presented in this thesis, articles have been published in the following areas: hashing-based learning for similarity search and recommendation \cite{DBLP:conf/www/hansen-www2021-phd, DBLP:conf/www/hansen-www2021-mish, DBLP:conf/sigir/hansen-sigir2020-coldstart, DBLP:conf/sigir/hansen-sigir2020-pairwisehash, DBLP:conf/sigir/hansen-sigir19-semhash}, text representation and classification \cite{dongsheng-ecir2021,MATIN2019199,hansen2018copenhagen,hansen2019copenhagen,DBLP:conf/sigir/hansen-sigir2019-contextprop,hansen-etal-2021-automatic}, educational datamining \cite{DBLP:conf/edm/edm2017,DBLP:conf/edm/hansen-edm-2019,ecel-2017}, and health-oriented modelling \cite{jimenez2021developing,DBLP:conf/cikm/hansen-cikm2017,trec-covid-lucas}. These articles are listed below:
\begin{itemize}
    \item Casper Hansen$^*$, Christian Hansen$^*$, Lucas Chaves Lima (2021). Automatic Fake News Detection: Are Models Learning to Reason? In ACL, pages 80-86. \cite{hansen-etal-2021-automatic}. 
    \item Christian Hansen$^*$, Casper Hansen$^*$, Jakob Grue Simonsen, Christina Lioma (2021). Projected Hamming Dissimilarity for Bit-Level Importance Coding in Collaborative Filtering. In WWW, pages 261-269. \cite{DBLP:conf/www/hansen-www2021-phd}.
    \item Christian Hansen$^*$, Casper Hansen$^*$, Jakob Grue Simonsen, Stephen Alstrup, Christina Lioma (2021). Unsupervised Multi-Index Semantic Hashing. In WWW, pages 2879-2889. \cite{DBLP:conf/www/hansen-www2021-mish}.
    \item Dongsheng Wang$^*$, Casper Hansen$^*$, Lucas Chaves Lima, Christian Hansen, Maria Maistro, Jakob Grue Simonsen, Christina Lioma (2021). Multi-Head Self-Attention with Role-Guided Masks. In ECIR, in press. \cite{dongsheng-ecir2021}.
    \item Espen Jimenez Solem, Tonny Studsgaard Petersen, Casper Hansen, Christian Hansen, et al. (2021). Developing and Validating COVID-19 Adverse Outcome Risk Prediction Models from a Bi-national European Cohort of 5594 Patients. In Scientific Reports 11 (1), pages 1-12. \cite{jimenez2021developing}.
    \item Lucas Chaves Lima$^*$, Casper Hansen$^*$, Christian Hansen, Dongsheng Wang, Maria Maistro, Birger Larsen, Jakob Grue Simonsen, Christina Lioma (2021). Denmark's Participation in the Search Engine TREC COVID-19 Challenge: Lessons Learned about Searching for Precise Biomedical Scientific Information on COVID-19. In TREC COVID-19 Challenge. \cite{trec-covid-lucas}.
    \item Casper Hansen$^*$, Christian Hansen$^*$, Jakob Grue Simonsen, Stephen Alstrup, Christina Lioma (2020). Content-aware Neural Hashing for Cold-start Recommendation. In SIGIR, pages 971-980. \cite{DBLP:conf/sigir/hansen-sigir2020-coldstart}.
    \item Casper Hansen$^*$, Christian Hansen$^*$, Jakob Grue Simonsen, Stephen Alstrup, Christina Lioma (2020). Unsupervised Semantic Hashing with Pairwise Reconstruction. In SIGIR, pages 2009-2012. \cite{DBLP:conf/sigir/hansen-sigir2020-pairwisehash}.
    \item Casper Hansen, Christian Hansen, Jakob Grue Simonsen, Christina Lioma (2019). Neural Weakly Supervised Fact Check-Worthiness Detection with Contrastive Sampling-Based Ranking Loss. In CLEF-2019 Fact Checking Lab. \cite{hansen2019copenhagen}.
    \item Casper Hansen, Christian Hansen, Stephen Alstrup, Jakob Grue Simonsen, Christina Lioma (2019). Contextually Propagated Term Weights for Document Representation. In SIGIR, pages 897-900. \cite{DBLP:conf/sigir/hansen-sigir2019-contextprop}.
    \item Casper Hansen, Christian Hansen, Jakob Grue Simonsen, Stephen Alstrup, Christina Lioma (2019). Unsupervised Neural Generative Semantic Hashing. In SIGIR, pages 735-744. \cite{DBLP:conf/sigir/hansen-sigir19-semhash}.
    \item Christian Hansen, Casper Hansen, Stephen Alstrup, Christina Lioma (2019). Modelling End-of-Session Actions in Educational Systems. In EDM, pages 306-311. \cite{DBLP:conf/edm/hansen-edm-2019}.
    \item Rastin Matin, Casper Hansen, Christian Hansen, Pia Mølgaard (2019). Predicting Distresses using Deep Learning of Text Segments in Annual Reports. In Expert Systems With Applications (132), pages 199-208. \cite{MATIN2019199}.
    \item Casper Hansen, Christian Hansen, Jakob Grue Simonsen, Christina Lioma (2018). The Copenhagen Team Participation in the Check-Worthiness Task of the Competition of Automatic Identification and Verification of Claims in Political Debates of the CLEF2018 CheckThat! Lab. In CLEF-2018 Fact Checking Lab. \cite{hansen2018copenhagen}.
    \item Casper Hansen, Christian Hansen, Stephen Alstrup, Christina Lioma (2017). Smart City Analytics: Ensemble-Learned Prediction of Citizen Home Care. In CIKM, pages 2095-2098. \cite{DBLP:conf/cikm/hansen-cikm2017}.
    \item Stephen Alstrup, Casper Hansen, Christian Hansen, Niklas Hjuler, Stephan Lorenzen, Ninh Pham (2017). DABAI: A data driven project for e-Learning in Denmark. In ECEL, pages 18-24. \cite{ecel-2017}.
    \item Christian Hansen, Casper Hansen, Niklas Hjuler, Stephen Alstrup, Christina Lioma (2017). Sequence Modelling For Analysing Student Interaction with Educational Systems. In EDM, pages 232-237. \cite{DBLP:conf/edm/edm2017}.
\end{itemize}

\chapter{Shifting Consumption towards Diverse Content on Music Streaming Platforms}\label{chapter1}
Christian Hansen, Rishabh Mehrotra, Casper Hansen, Brian Brost, Lucas Maystre, Mounia Lalmas (2021). Shifting Consumption towards Diverse Content on Music Streaming Platforms. In WSDM, pages 238-246. \cite{wsdm-spotify}.
\includepdf[pages=-, pagecommand={}]{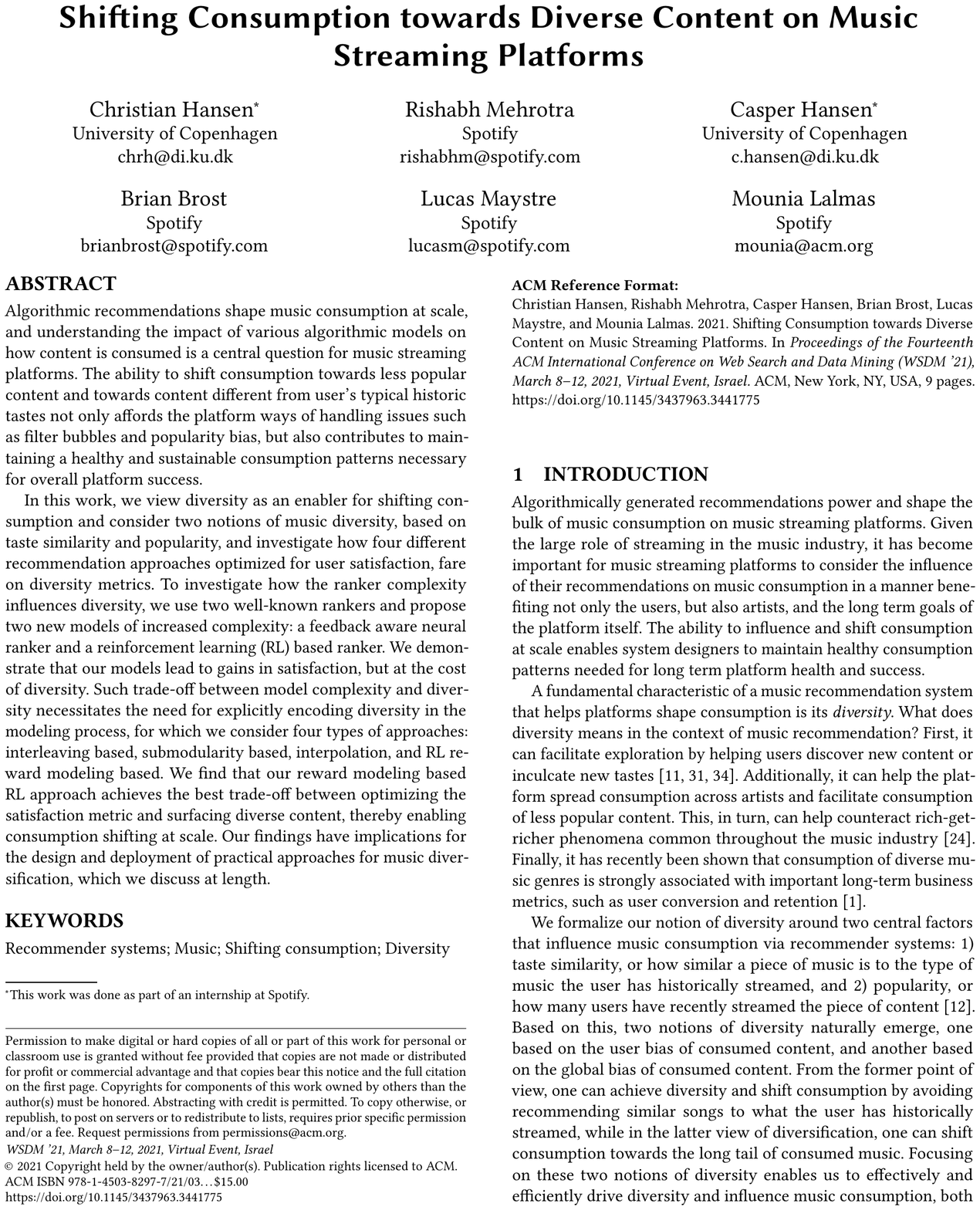}

\chapter{Contextual and Sequential User Embeddings for Large-Scale Music Recommendation}\label{chapter2}
Casper Hansen, Christian Hansen, Lucas Maystre, Rishabh Mehrotra, Brian Brost, Federico Tomasi, Mounia Lalmas (2020). Contextual and Sequential User Embeddings for Large-Scale Music Recommendation. In RecSys, pages 53-62. \cite{DBLP:conf/recsys/hansen-recsys-spotify}.
\includepdf[pages=-, pagecommand={}]{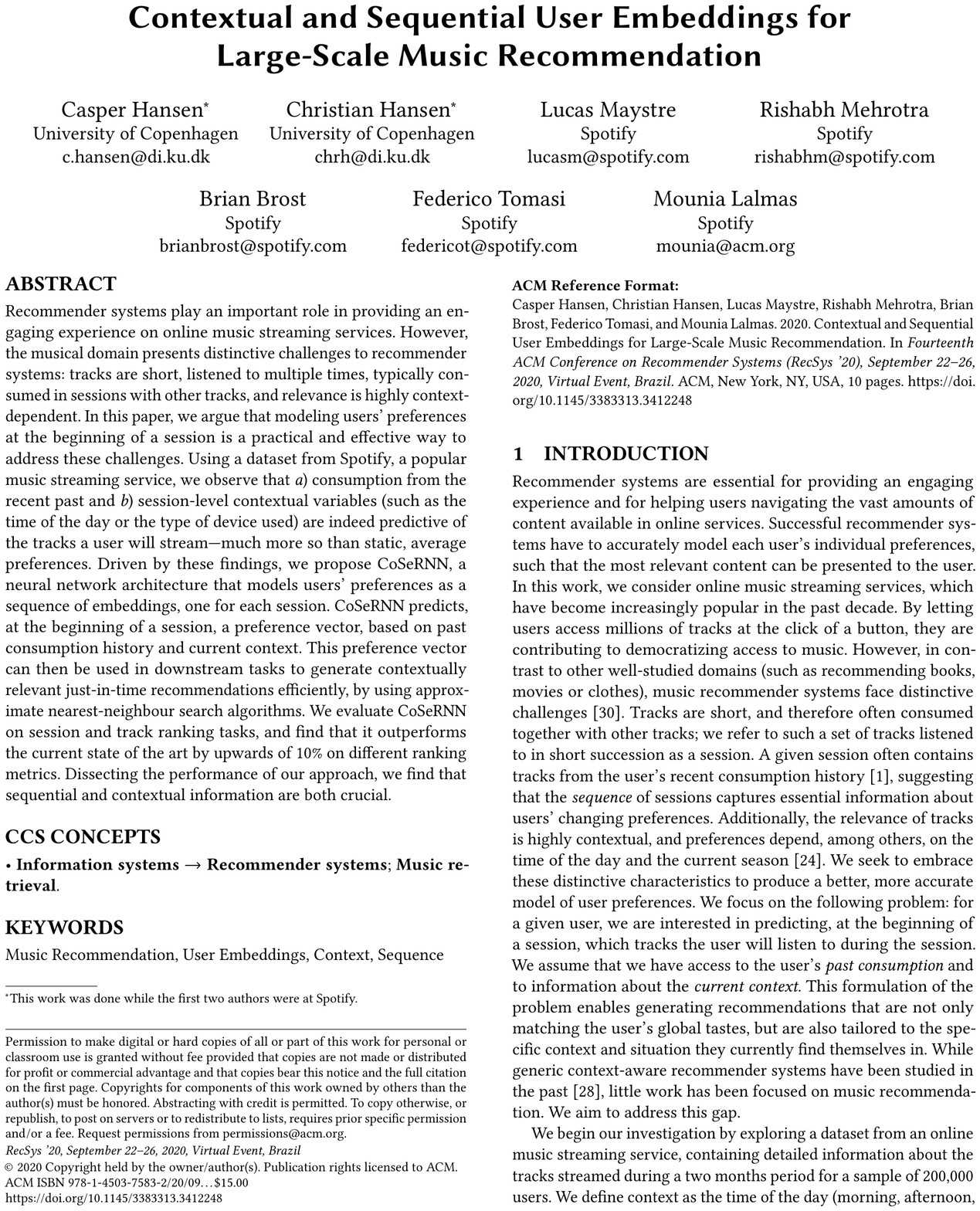}

\chapter{Modelling Sequential Music Track Skips using a Multi-RNN Approach}\label{chapter3}
Christian Hansen, Casper Hansen, Jakob Grue Simonsen, Stephen Alstrup, Christina Lioma (2019). Modelling Sequential Music Track Skips Using a Multi-RNN Approach. In WSDM Cup. \cite{DBLP:conf/www/hansen-wsdmcup-2019}.
\includepdf[pages=-, pagecommand={}]{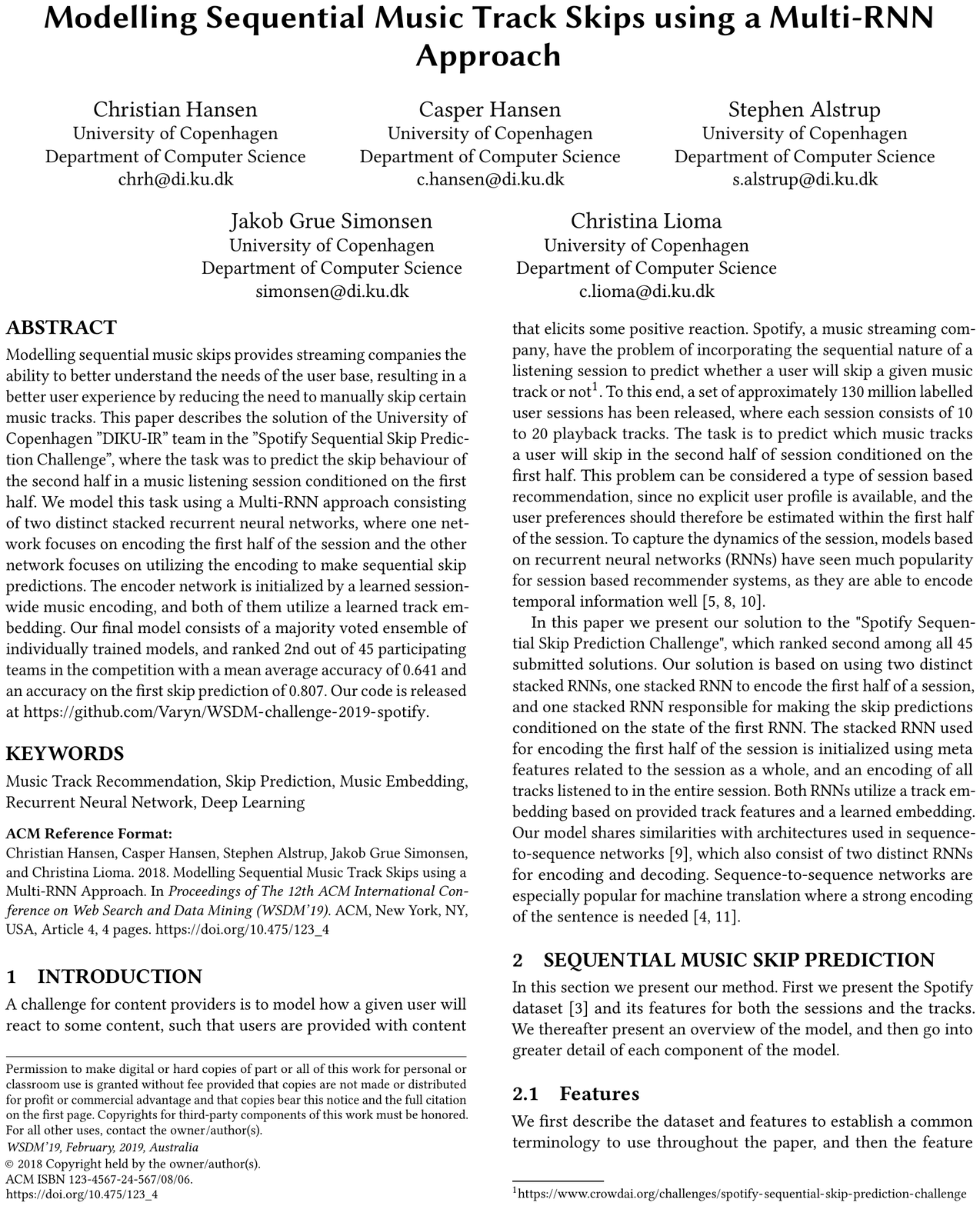}

\chapter{MultiFC: A Real-World Multi-Domain Dataset for Evidence-Based Fact Checking of Claims}\label{chapter4}
Isabelle Augenstein, Christina Lioma, Dongsheng Wang, Lucas Chaves Lima, Casper Hansen, Christian Hansen, Jakob Grue Simonsen (2019). MultiFC: A Real-World Multi-Domain Dataset for Evidence-Based Fact Checking of Claims. In EMNLP, pages 4685-4697. \cite{augenstein-etal-2019-multifc}.
\includepdf[pages=-, pagecommand={}]{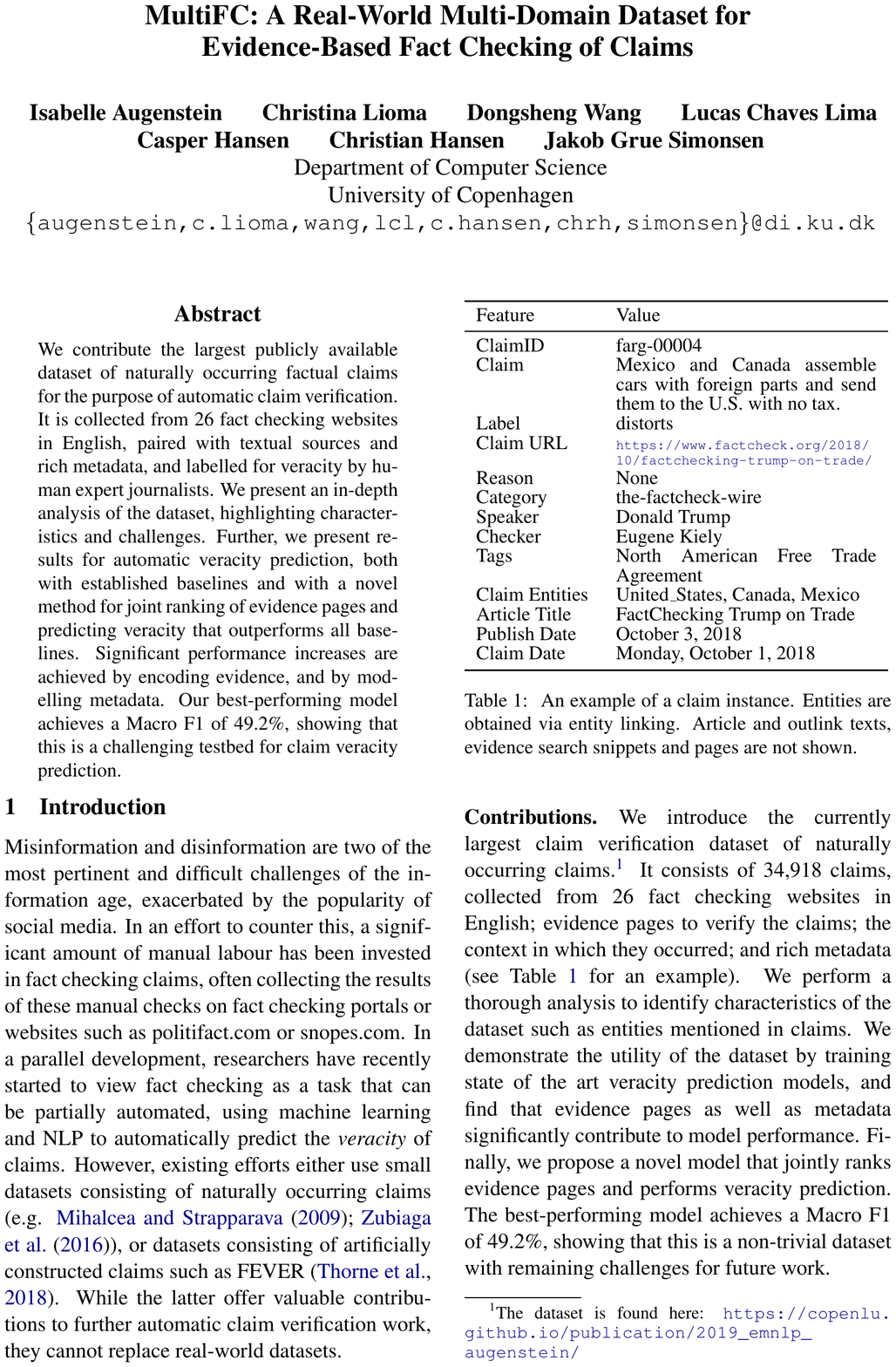}

\chapter{Factuality Checking in News Headlines with Eye Tracking}\label{chapter5}
Christian Hansen, Casper Hansen, Jakob Grue Simonsen, Birger Larsen, Stephen Alstrup, Christina Lioma (2020). Factuality Checking in News Headlines with Eye Tracking. In SIGIR, pages 2013-2016. \cite{DBLP:conf/sigir/sigir2020-eyetracking}.
\includepdf[pages=-, pagecommand={}]{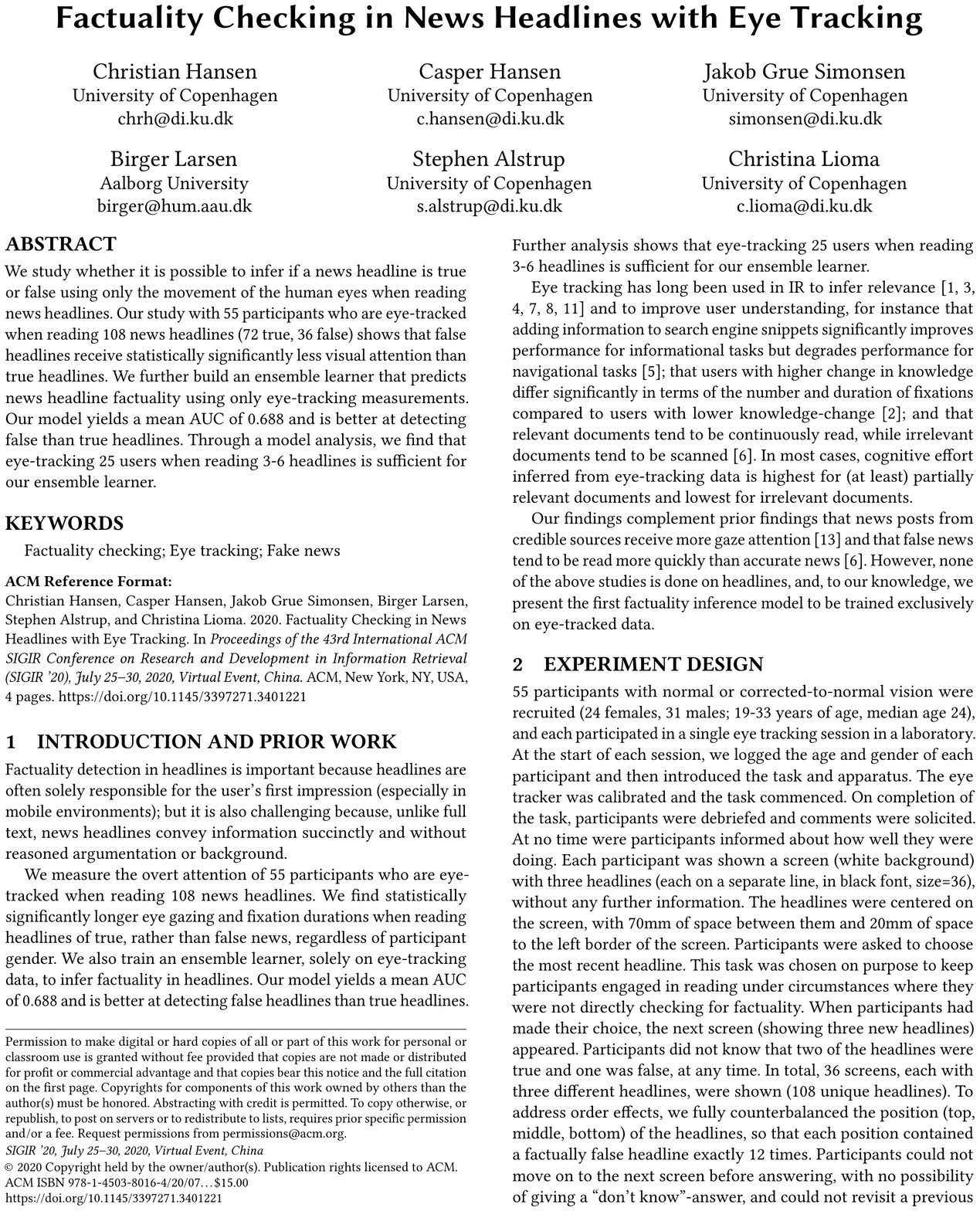}

\chapter{Neural Check-Worthiness Ranking with Weak Supervision: Finding Sentences for Fact-Checking}\label{chapter6}
Casper Hansen, Christian Hansen, Stephen Alstrup, Jakob Grue Simonsen, Christina Lioma (2019). Neural Check-Worthiness Ranking with Weak Supervision: Finding Sentences for Fact-Checking. In Companion Proceedings of WWW, pages 994-1000. \cite{DBLP:conf/www/hansen-www2019-factcheck}.
\includepdf[pages=-, pagecommand={}]{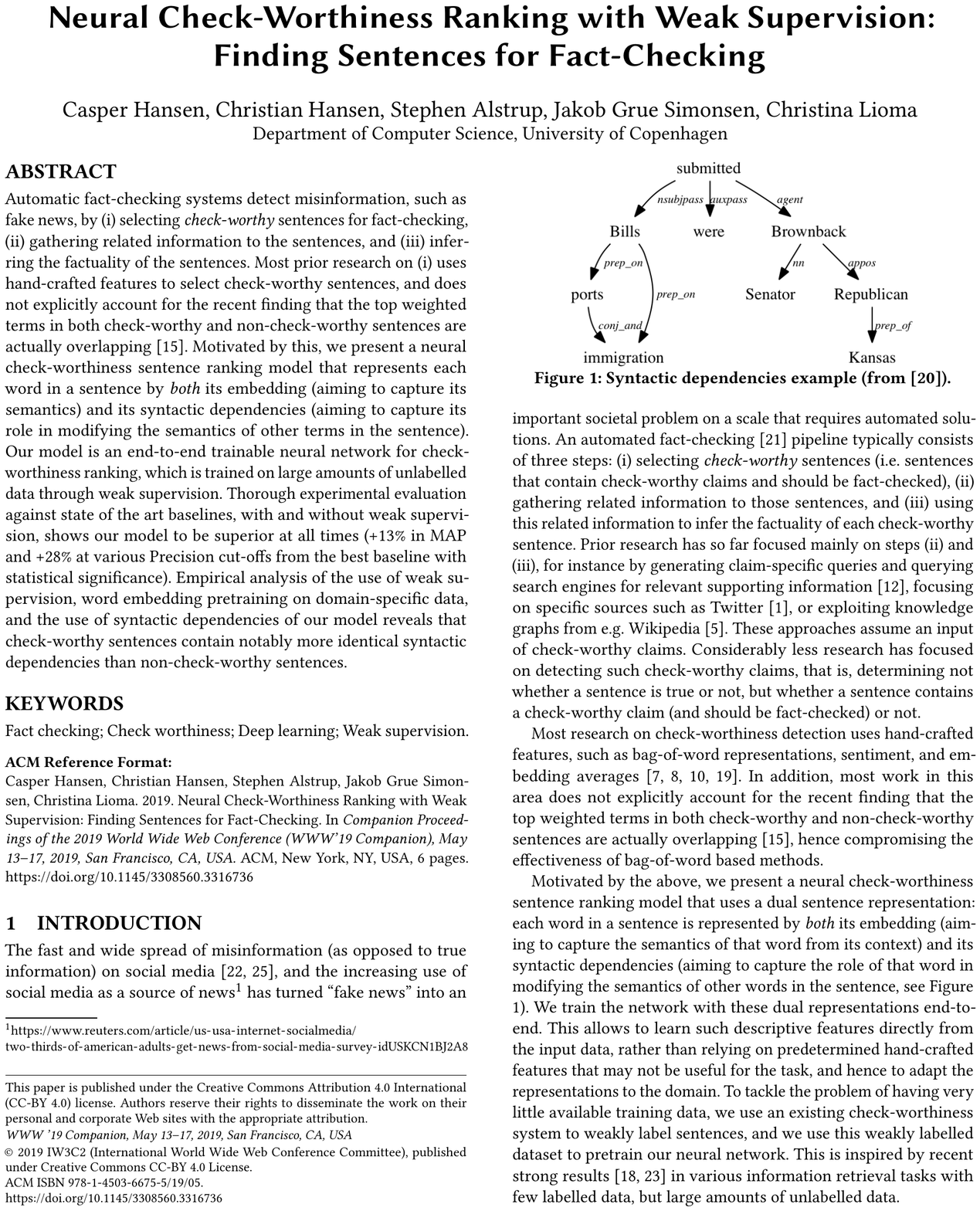}

\chapter{Fact Check-Worthiness Detection with Contrastive Ranking}\label{chapter7}
Casper Hansen, Christian Hansen, Jakob Grue Simonsen, Christina Lioma (2020). Fact Check-Worthiness Detection with Contrastive Ranking. In CLEF, pages 124-130. \cite{DBLP:conf/clef/hansen-clef-2020}.
\includepdf[pages=-, pagecommand={}]{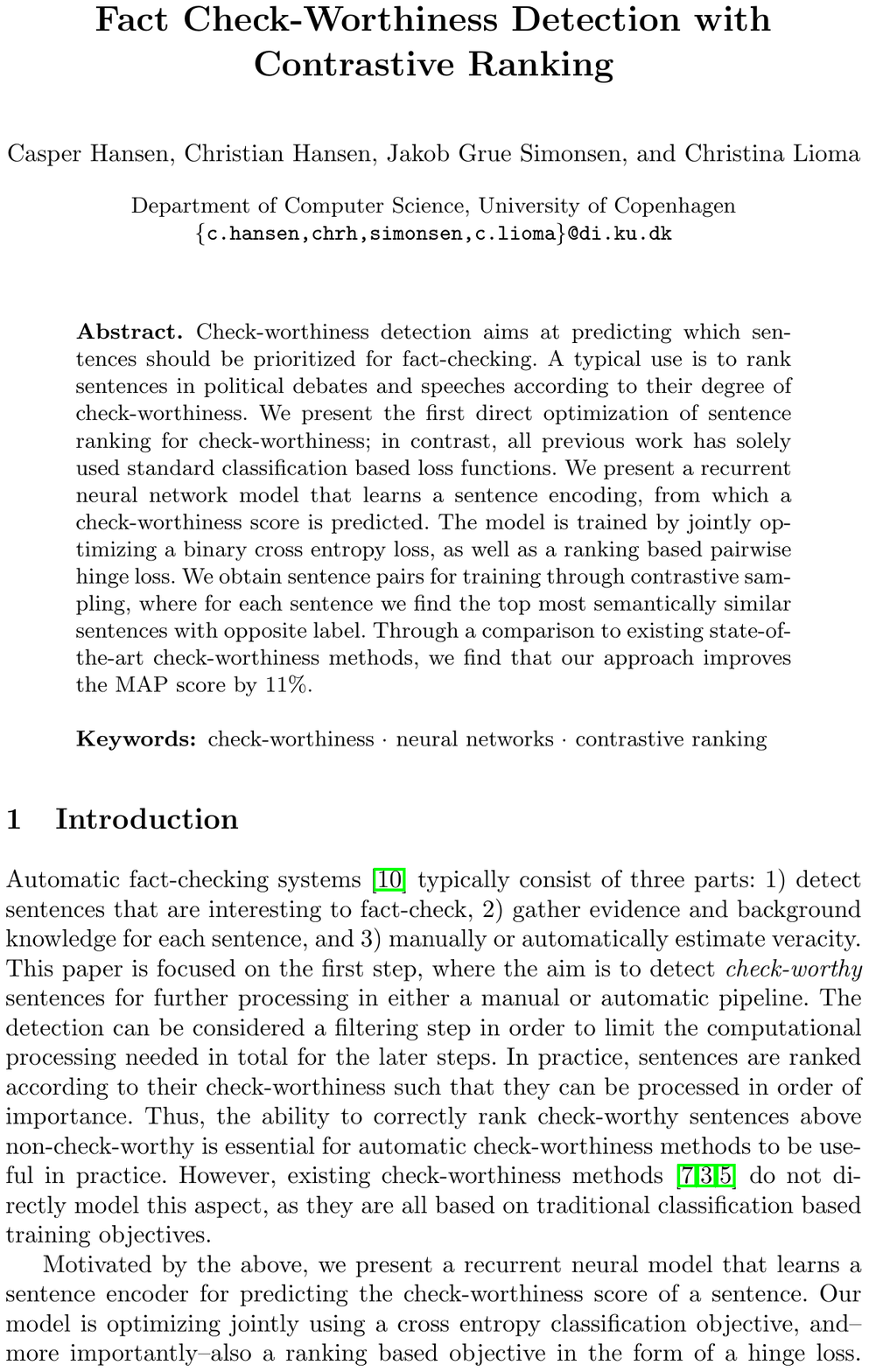}

\chapter{Neural Speed Reading with Structural-Jump-LSTM}\label{chapter8}
Christian Hansen, Casper Hansen, Stephen Alstrup, Jakob Grue Simonsen, Christina Lioma (2019). Neural Speed Reading with Structural-Jump-LSTM. In ICLR. \cite{DBLP:conf/iclr/hansen-iclr}.
\includepdf[pages=-, pagecommand={}]{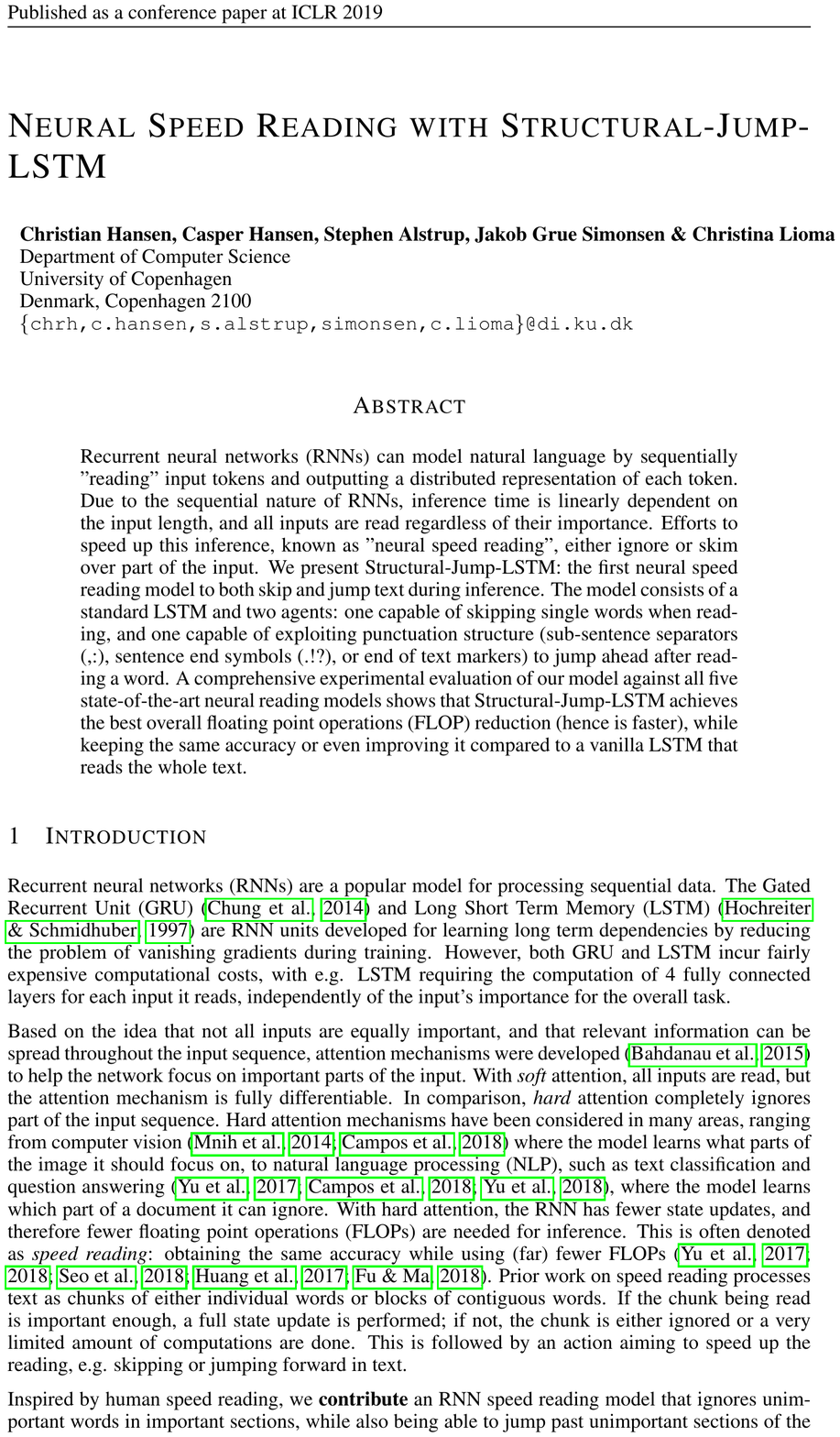}

\phantomsection
\addcontentsline{toc}{chapter}{Bibliography}
\bibliographystyle{plain}
\bibliography{references}

\end{document}